# Genome Sequencing Highlights Genes Under Selection and the Dynamic Early History of Dogs

Adam H. Freedman[1], Rena M. Schweizer[1], Ilan Gronau[2], Eunjung Han[1], Diego Ortega-Del Vecchyo[1], Pedro M. Silva[3], Marco Galaverni[4], Zhenxin Fan[5], Peter Marx[6], Belen Lorente-Galdos[7], Holly Beale[8], Oscar Ramirez[7], Farhad Hormozdiari[9], Can Alkan[10], Carles Vilà[11], Kevin Squire[12], Eli Geffen[13], Josip Kusak[14], Adam R. Boyko[15], Heidi G. Parker[8], Clarence Lee[16], Vasisht Tadigotla[16], Adam Siepel[2], Carlos D. Bustamante[17], Timothy T. Harkins[16], Stanley F. Nelson[12], Elaine A. Ostrander[8], Tomas Marques-Bonet[7,18], Robert K. Wayne[1*], John Novembre[1,19*]

[1]Department of Ecology and Evolutionary Biology, University of California, Los Angeles, Los Angeles, CA, USA
[2]Department of Biological Statistics and Computational Biology, Cornell University, Ithaca, NY, USA
[3]CIBIO-UP, University of Porto, Vairão, Portugal
[4]ISPRA, Ozzano dell'Emilia, Italy
[5] Key Laboratory of Bioresources and Ecoenvironment, Sichuan University, Chengdu, China
[6]Department of Measurement and Information Systems, Budapest University of Technology and Economics, Budapest, Hungary
[7]Institut de Biologia Evolutiva (CSIC-Univ Pompeu Fabra), Barcelona, Spain
[8]National Institutes of Health/NHGRI, Bethesda, MD, USA
[9]Department of Computer Science, University of California, Los Angeles, Los Angeles, CA, USA
[10]Bilkent University, Ankara, Turkey
[11]Estación Biológia de Doñana EBD-CSIC, Sevilla, Spain
[12]Department of Human Genetics, University of California, Los Angeles, Los Angeles, CA, USA
[13]Department of Zoology, Tel Aviv University, Tel Aviv, Israel
[14]University of Zagreb, Zagreb, Croatia
[15]Department of Veterinary Medicine, Cornell University, Ithaca, NY
[16]Life Technologies, Foster City, CA, USA
[17]Stanford School of Medicine, Stanford, CA, USA
[18]Institució Catalana de Recerca i Estudis Avançats (ICREA). 08010, Barcelona, Spain
[19]Current address: Department of Human Genetics, University of Chicago, Chicago, IL,USA

*Correspondence to: jnovembre@uchicago.edu

## Abstract

To identify genetic changes underlying dog domestication and reconstruct their early evolutionary history, we analyzed novel high-quality genome sequences of three gray wolves, one from each of three putative centers of dog domestication, two ancient dog lineages (Basenji and Dingo) and a golden jackal as an outgroup. We find dogs and wolves diverged through a dynamic process involving population bottlenecks in both lineages and post-divergence gene flow, which confounds previous inferences of dog origins. In dogs, the domestication bottleneck was severe involving a 17 to 49-fold reduction in population size, a much stronger bottleneck than estimated previously from less intensive sequencing efforts. A sharp bottleneck in wolves occurred soon after their divergence from dogs, implying that the pool of diversity from which dogs arose was far larger than represented by modern wolf populations. Conditional on mutation rate, we narrow the plausible range for the date of initial dog domestication to an interval from 11 to 16 thousand years ago. This period predates the rise of agriculture and, along with new evidence from variation in amylase copy number, implies that the earliest dogs arose alongside hunter-gathers rather than agriculturists. Regarding the geographic origin of dogs, we find that surprisingly, none of the extant wolf lineages from putative domestication centers are more closely related to dogs, and the sampled wolves instead form a sister monophyletic clade. This result, in combination with our finding of dog-wolf admixture during the process of domestication, suggests a re-evaluation of past hypotheses of dog origin is necessary. Finally, we also detect signatures of selection, including evidence for selection on genes implicated in morphology, metabolism, and neural development. Uniquely, we find support for selective sweeps at regulatory sites suggesting gene regulatory changes played a critical role in dog domestication.

## Introduction

Historically, gray wolves have been dominant predators across Eurasia and North America, often exerting top-down impacts on the ecological communities they inhabit [1,2]. As humans expanded out of Africa into Eurasia, they came into contact with gray wolves and, through a complex and poorly understood process, dogs emerged as the first human companion species and the only large carnivore ever to be domesticated. Archaeological evidence provides partial clues about dog origins. For example, dog-like canids first appear in the fossil record as early as 33,000 years ago in Siberia [3]. However, it is not clear if these proto-dog fossils represent failed domestication attempts or are ancestral to modern dogs. Similarly, the geographic origin of dogs is uncertain, with different lines of evidence supporting Southeast Asia, the Middle East, and Europe as potential domestication centers, and ruling out Africa, Australia, and North America [4-10]. Nonetheless, the genetic basis of several traits that changed during dog domestication and breed formation are becoming increasingly illuminated, and these studies advance the general understanding of how genetic mechanisms shape phenotypic trait diversity [11-14]. For example, a recent study found an increase in amylase copy number during dog domestication suggesting adaptation to starch rich diets [15]. Given the unique behavioral adaptations of dogs, including docility and the ability to form social bonds with humans [16], comparative genomics analyses of dogs and wolves holds great promise for identifying genetic loci involved in complex behavioral traits [14].

Here, we analyze 10 million single-nucleotide variant sites from whole-genome data we generated for six unique canid lineages to advance the understanding of dog origins and genetic changes early in dog domestication. These data include whole-genome sequences of three individual wolves (*Canis lupus*), an Australian Dingo, a Basenji and a golden jackal (*Canis aureus*). With these data, we investigated: 1) the size of the ancestral wolf population at the time of wolf/dog divergence; 2) the geographic origins and timing of dog domestication; 3) post-divergence admixture between dogs and wolves; 4) the types of variation (regulatory versus structural) most strongly selected during domestication; 5) specific loci that underwent positive selection during domestication; and 6) lineage-specific characteristics of the recently discovered dog-specific amylase expansion [15].

## Results

### Individual-level genome sequences.

The three wolves sequenced were chosen to represent the broad regions of Eurasia where domestication is hypothesized to have taken place (Europe, the Middle East, and East/Southeast Asia) [5], and specifically were sampled from Croatia, Israel, and China (Fig. 1A). Further, we sampled the Dingo and Basenji because relative to the reference Boxer genome, they are divergent lineages [7] and maximize the odds to capture distinct alleles present in the earliest dogs. These lineages are also geographically distinct, with modern Basenjis tracing their history to hunting dogs of western Africa, while Dingoes are free-living semi-feral dogs of Australia that arrived there at least 3500 years ago (Fig. 1A) [17]. As a result of their geographic isolation these two dog lineages are less likely to have overlapped with and admixed with wolves in the recent past. Sequencing the golden jackal allowed us to identify the ancestral state of variants arising in dogs and wolves (Text S1-S2). For some analyses, we also leverage data from a companion study of 12 additional dog breeds (Text S1).

For each of the six samples, we generated high-quality genome sequences and this approach enables individual-level analyses, such as the reconstruction of lineage-specific demographic histories. Cumulative coverage was 72x for the wolves (24x average per individual), 38x coverage for the two dogs (19x average per individual), and 24x for the golden jackal, for a total of 335Gb of uniquely aligned sequence from 11.2 billion reads (Table S2.2). Surveys of wolf genetic diversity to date have been limited to shotgun sequencing with incomplete genomic coverage [18], small numbers of sequence loci [19], limited pooled sequencing (6x average from a pool of 12 wolves, 30x average from a pool of 60 dogs) [15] or lower coverage sequencing (9-11x coverage of 4 wolves, 9-14x of 7 dogs) [20]. We chose a design based on higher coverage of a small number of dog/wolf genomes and an outgroup genome, because recent advances have made it possible to use small (and even single) samples to gain extensive information about past demography [21-23], and loci that have undergone positive selection have been found in other contexts using small samples in a comparative framework [24,25].

Our analyses draw on 10,265,254 high quality variants detected by our genotyping pipeline (Text S3-S5), of which 6,970,672 were at genomic positions with no missing data for any lineage (Table S5.1.1-S5.1.2). We estimate genotype error rates to be very low based on comparison to genotype calls from genotyping arrays (e.g.

heterozygote discordance rates of 0.01-0.04%, Table S5.2.1-S5.2.2, Text S5). Further, PCA on the intersection of the sequencing variants and genotyping array variants show the novel samples cluster appropriately, suggesting batch effects due to technology have been minimized (Fig. 1C, Text S5).

**Demographic bottlenecks in dogs and wolves.**

Genome-wide patterns of heterozygosity provide an indication of long-term effective population sizes. We observed mean genome-wide heterozygosity rates per nucleotide of 0.09% for the Basenji, and 0.06% for the Dingo, lower than those for gray wolves (0.12%-0.16%, Fig. 1B, Table S5.6.1), and consistent with a domestication bottleneck. Similarly low heterozygosity rates of 0.06% were previously observed in modern dog breeds [18]. To better understand the differences in the demographic histories of dogs and wolves that cause this two-fold difference in heterozygosity, we inferred trajectories of ancestral effective population sizes ($N_e$) using the pairwise sequential Markovian coalescent (PSMC) method (Text 9.2) [22]. The results show similar inferred population sizes for all five genomes until the trajectories appear to diverge about 1.7 x $10^{-4}$ substitution units in the past (i.e. 50kya, assuming an average mutation rate, $\mu$, of $1\times10^{-8}$ per generation and three years per generation [18]). The ancestral $N_e$ of dogs appears to decline from roughly 35,000 individuals (Basenji: 32,100-35,500; Dingo: 32,500-37,400 95% bootstrap CI) at the divergence point to lineage-specific estimates that fall below 2,000 (Basenji: 1640-1980, 95% CI at 4 kya; Dingo: 704-1042 95%CI at 3 kya), implying a domestication bottleneck equivalent to a 17 to 49-fold reduction in effective population size. After the divergence from dogs, all three wolf populations show a slight increase, followed by a notable decline in $N_e$ continuing until the present value of 10,000 to 17,000 individuals. These declines do not appear to be due to very recent inbreeding as we found runs of homozygosity do not affect our inferences of ancestral $N_e$ (Text 9.2.2). Thus, the observation of a relatively minor two-fold difference in current diversity between dogs and wolves is explained by paired bottlenecks, with that of wolves being less extreme than dogs. One implication of our result is that previous studies that have not modeled the wolf bottleneck have underestimated the bottleneck associated with dog domestication [18,19].

**Recent sequence divergence**

To place patterns of diversity and demographic trends in a phylogenetic context, and to identify a starting topology for building a comprehensive model of canid demographic history, we used pairwise sequence divergence computed from the genome-wide data to construct a neighbor joining (NJ) tree for the six sequenced genomes and the boxer reference, with the golden jackal as an outgroup (Fig. S9.1.1). The NJ tree shows dogs and wolves as monophyletic sister clades, and surprisingly shows the two dog genomes being only slightly more similar to the Boxer reference than the three wolf genomes. Given the short internal branches of the NJ tree and small range of pairwise divergences within the six dogs and wolves (0.087%– 0.118%, see Table S.9.1.2), we expect incomplete lineage sorting and ancestral polymorphisms to contribute considerably to the patterns of variation in dogs and wolves. Indeed, of the variant sites with no missing data, sites with shared variation across the dogs and wolves greatly outnumber the variants that are private to the dogs or the wolves (22% of variants shared; 6% private to dogs; 12% private to wolves), and only 0.3% of variants were fixed between the dog and wolf samples (Table S5.1.2, see also Text S9). These signatures are consistent with incomplete lineage sorting due to recent divergence, but could also be caused by post-divergence gene flow/admixture between dogs and wolves.

**Admixture between wolves and dogs**

To assess admixture, we employed the 'ABBA-BABA' test, which was developed to detect gene flow between two divergent populations, such as humans and Neandertals [24] and uses individual genome sequences and a non-parametric statistical approach (Text S9). We applied this test to all pairs of samples using the golden jackal as outgroup, and found significant evidence of geographically consistent post-divergence gene flow, between Israeli wolf and Basenji ($\hat{Z}$ = 9.11), Chinese wolf and Dingo ($\hat{Z}$ = 6.20), and Israeli wolf and Boxer ($\hat{Z}$ = 6.06) (Table S9.4.1). The signal for Chinese wolf and Dingo likely represents admixture in Eastern Eurasia between populations of dogs and wolves ancestral to the Chinese wolf and Dingo. Similarly, the signal observed for Israeli wolf, Basenji, and Boxer likely represents ancestral admixture that occurred in western Eurasia (see discussion below).

**Inference of a demographic model of dog domestication**.

We next constructed a unified demographic model for dogs and wolves, with estimates of population divergence times, ancestral population sizes and rates of post-divergence gene flow, by jointly analyzing all genomes using a recently developed Bayesian coalescent-based method (Generalized Phylogenetic Coalescent Sampler; *G-PhoCS*) [21]. *G-PhoCS* produces estimates of demographic parameters by extracting patterns of variation in genealogies at neutrally evolving loci genome-wide, using a full probabilistic model that accounts for both incomplete lineage sorting and post-divergence gene flow.

Applying *G-PhoCS* to the six genomes and Boxer reference using the population phylogeny suggested in Fig. S9.1.1, we infer that dogs and wolves diverged 0.5 x $10^{-4}$ substitution units ago with 0.45-0.53 x $10^{-4}$ 95% Bayesian credible interval. This corresponds to a divergence time of 15 kya with a 13.9–15.9 kya CI using the calibration of $\mu$=1 x $10^{-8}$/generation; 3 years/gen, Fig. 2A. Further, *G-PhoCS* estimates significant rates of post-divergence gene flow between Israeli wolf and Basenji and between Chinese wolf and Dingo, consistent with the non-parametric ABBA/BABA tests. We also inferred gene flow between the golden jackal and the Israeli wolf, and between the jackal and the dog-wolf ancestor, that could not be tested using the ABBA/BABA test. While we also detected admixture between Israeli wolf and Boxer, this signal is likely a result of gene flow from Basenji to Israeli wolf (Text S10).

We also considered alternative topologies to the population phylogeny given in Fig 2A. While *G-PhoCS* does not yet support statistical tests of phylogenetic relationships, simulations indicate *G-PhoCS* frequently infers high levels of gene flow or compresses divergence events together when an incorrect tree topology is assumed. When we fit models with unique domestications of the Boxer, Basenji, and Dingo (e.g, Fig 2B), the inferred levels of post-domestication gene flow are large and unrealistic (e.g. Basenji-Dingo m=0.47 CI:0.29-0.66, full results Fig. S10.6). When we fit a model in which the three dogs have a single origin from one of the sampled wolf lineages (e.g., Fig. 2C), the resulting divergences between all wolf and dog populations become temporally compressed, also suggesting poor model fit (full results Fig. S10.7). A key observation from considering alternative phylogenies (Text S10) is that, regardless of which phylogeny is assumed, the underlying sequence divergences and resulting parameter inferences suggest all three sampled wolves are equally divergent from the dogs.

To further investigate robustness of the *G-PhoCS* estimation, we confirmed via simulation that the method produces accurate estimates of divergence times regardless of whether post-divergence bottlenecks took place gradually or abruptly (Text S10). We also find the apparent discrepancy between the population divergence time inferred by *G-PhoCS* ($0.5 \times 10^{-4}$ substitution units) and that implied by the PSMC results ($\sim 1.7 \times 10^{-4}$ substitution units) (Fig. 1D) appears to be due to the behavior of PSMC. Under a model with an abrupt change in $N_e$, such as likely occurs in domestication, the PSMC method infers gradual changes in $N_e$, and this pushes backwards the apparent time of divergence among trajectories (Fig. S9.2.3 – S9.2.5). We additionally considered several alternative sets of loci in the analysis and found little impact on the results (Text S10). Finally, although the timing of divergence between the golden jackal and the dog-wolf ancestor was more recent than previously reported [26], comparison of polymorphism data in our jackal genome to a larger panel of wolf and jackals validates its positions within the golden jackal lineage (Text S5.4.3, S16).

**Regulatory evolution on the dog lineage**.

Strong selection favoring a novel advantageous mutation will reduce diversity in neutral regions flanking the selected site, in a process known as a selective sweep [27]. To compare the strength of selective sweeps across site categories, we identified sites where derived alleles in dogs are at high frequency or are completely fixed, then stratified these sites by functional class, and evaluated the extent to which diversity was reduced around such sites (Text S12) [28,29]. Remarkably, we found the strongest reductions in diversity near 5' ($p<0.0002$) and 3' UTR ($p=0.012$) sites (Fig. 3). To a lesser extent we observed sweep signatures at conserved non-coding (CNE) elements ($p=0.037$; identified using phastCons scores derived from an alignment of multiple mammalian genomes, Text S7). The diversity around non-synonymous sites is not reduced, and actually appears inflated ($p=0.0014$). These results contrast to those in humans, where only weak evidence of selective sweeps is found using similar methods [28]. Our results suggest that regulatory variants (UTR) and variants within CNEs (that may comprise regulatory elements) have experienced stronger sweeps than structural, nonsynonymous variants during dog domestication.

**Regions under selection during domestication**

Identifying genes under recent positive selection is inherently challenging [30,31], but our sampling provides the opportunity to identify loci potentially under selection on the dog lineage during domestication. First, we scanned the autosomal genome for signatures of positive selection on the dog lineage using three metrics ($F_{ST}$, $\Delta\pi$, and $\Delta$Tajima's D) that have been shown to have high power to detect regions under selection during domestication [32,33]. We flagged extreme outliers in 100kb windows based on a joint percentile of these metrics, and then identified clusters of outliers to establish candidate selection regions (see Text S13.1 for details).

The top 100 outlier regions range in length from 10-530kb (Fig. 3B, S13.1-S13.7). Forty-five of the top 100 regions did not contain any validated, annotated genes. These regions may harbor important non-coding functional elements (CNEs) but might also include unidentified coding regions. In support of the former, we observe a 1.6-fold enrichment in CNEs in these regions relative to the genome-wide distribution (2.5% in outlier regions without genes vs. 1.6% genome-wide, Fisher's exact test, $p=2.2 \times 10^{-16}$). Several of the genes in our top regions overlap with previous studies or with a re-analysis of previous SNP array data, in which we contrasted variation between wolves and basal dogs (Fig. 4B, Text S14).

We also assessed whether genes in the top 10% of our selection scan regions are enriched for particular functional groups of genes using Gene Ontology, Human Phenotype, and KEGG Pathway functional categories (see Text S13.2). Functional enrichments were dominated by the Human Phenotype categories, especially high-level morphological categories such as `abnormality of dental morphology' and 'abnormality of the joints of the upper limbs'. More specific categorizations were found as well (Table S13.1), including the category 'abnormality of the 5th finger' which we speculate could relate to the development of the dewclaw in dogs, which is absent in wild canids.

Several significant enrichment categories included genes involved in brain structural development (e.g. 'cerebellar malformation, 'cerebellar vermis hypoplasia', 'delayed closure of fontanelles') (Table S13.1). We also see impacts on neurological functions reflected among our very top candidate regions. Four of the eight top candidate regions contained genes implicated with neurological functions in other mammalian species: *CADM2* (under the 4$^{th}$ most extreme outlier region) is a synaptic cell adhesion molecule whose flanking regions show reduced homozygosity in autism patients [34]; *SH3GL2* (6$^{th}$ region) affects synaptic vesicle formation [35]; *PDE4D (7$^{th}$* region) is a mammalian homolog of the *dunce* gene in Drosophila [36] whose knockout in mice shows impaired learning [37]; and *CUX2* (8$^{th}$ region) is a key marker of neuronal fate during mammalian cortex development [38] whose knockout in mice shows deficits in working memory [39].

Focusing on our top outlier region for positive selection, we find it contains a portion of the *ELF2* gene at its flank, but the putative selection signature is most strongly peaked on *CCRN4L* (Fig. 4A). We find additional support for this signature in whole genome data from an additional 12 breeds (Fig. S13.5), and across 912 dogs from 85 breeds genotyped on a SNP array [7] (Fig. S13.6). *CCRN4L* (also known as Nocturnin) is expressed in a circadian fashion and studies in mice indicate *CCRN4L* activates *PPAR-γ*, a gene that promotes bone adipogensis as opposed to osteoblast formation and that harbors a known diabetes risk variant in humans [40]. It also is known to regulate the expression of genes involved in lipogenesis and fatty acid binding, and knock-out mice are remarkable in being resistant to diet-induced obesity [40-43]. *CCRN4L* also suppresses *IGF1,* a well-known activator of bone growth [42] that underlies size variation amongst dog breeds [44,45]. The direction of these pleiotropic effects of *CCR4NL* imply a gain-of-function mutation would promote adipocytes formation, alter lipid metabolism, and suppress bone-growth.

Another major locus of interest in dog-wolf domestication is the amylase gene *AMY2B,* as increase in copy number of the gene in dogs was recently identified as an adaptation that allowed early dogs to exploit a starch-rich diet as they fed on refuse from human agriculture [15]. In that study, copy number segregated between species, with only 2 copies in each of the 35 wolves genotyped and an average 7.4-fold increase across 136 dogs. Surprisingly, we find the Dingo has just 2 copies of amylase (Fig. 5A, Text S6), suggesting that amylase copy number expansion was not fixed across all dogs early in the process of domestication. In a survey of sequence data from 12 additional breeds, we find that, the Siberian Husky, a breed historically associated with nomadic hunter gatherers of the Arctic, has only 3-4 copies whereas the Saluki, which was historically bred in the Fertile Crescent where agriculture originated, has 29 copies (Fig. S15.1). Thus, amylase copy number corresponds to the degree to which breeds associated with agrarian human societies. In order to validate these results, we used real-time quantitative PCR (qPCR) to explore the variation in *AMY2B* copies across additional breed dogs (n=52), additional Dingoes (n=6) and a worldwide distribution of wolves (n=40) (Text S6.3). The qPCR results show modern dog breeds on average have high copy number of amylase and that wolves and Dingoes do not (Fig. 5B, Table S6.3.1). However, the qPCR also shows that amylase duplications are polymorphic in wolves (16 of 40 wolves Fig 5B) and thus are not restricted to dogs.

## Discussion

Using an array of analyses applied to high-quality individual canid genomes, we have examined the history of dogs and gray wolves in greater detail than previous studies. We demonstrate our sequence data are of high quality and we present new insights on the controversial question of the geography of dog domestication.

First, we found evidence of wolf-dog admixture in two dog lineages that have been isolated from wolves geographically in the recent past. This suggests wolf-dog admixture was occurring ancestrally and has impacted multiple, if not most, dog lineages [46,47]. Admixture has likely complicated previous inferences of dog origins. For instance, the presence of long shared haplotypes in Middle East wolves with several dog breeds [7] may reflect historic admixture rather than recent divergence. Similarly, higher genetic diversity in East Asian dogs and affinities between East Asian village dogs and wolves [4,6,20] may be confounded by past admixture with wolves. In areas where village dogs [48] roam freely and wolves have historically been in close proximity, admixture may also be present and have non-trivial impact on patterns of genetic variation.

Second, our demographic modeling with *G-PhoCS* supports best a population phylogeny in which dogs arise from a single domestication. Alternative models in which dog lineages arise from distinct wolf lineages are consistent with the data only if unrealistically high levels of gene flow are assumed between dogs or wolves, or divergences between lineages are nearly instantaneous (Figs. S10.6, S10.7). Surprisingly, no one wolf lineage of those we sampled was more closely related to dogs than any other, suggesting that the nearest wild relative to dogs was a more basal wolf than sampled in this study. Such a wolf lineage may exist today and not be represented by our samples; however, we consider this unlikely as we sampled the three major putative domestication regions, and previous SNP array studies have shown that wolf populations are only weakly differentiated, and so our sampled wolves should serve as good proxies for wolves in each broad geographic region [7]. One alternative is that the basal wolf lineage has gone extinct and the current wolf diversity from each region represents novel younger wolf lineages, as suggested by their recent divergence (Figs. 2A, S9.1.1). Our inference that wolves have gone through bottlenecks across Eurasia (Fig. 1D) suggests a dynamic period for wolf populations over the last 20,000 years and that extinction of particular lineages is not inconceivable. Indeed, several external lines of evidence provide support for substantial turnover in wolf lineages. For example, ancient DNA, isotope, and morphologic evidence identify a divergent North American Late Pleistocene wolf [49] and in Eurasia, similarly distinct wolves exist in the early archaeological record in Northern Europe and Russia, 15-36kya [8,9,50]. Presumed changes in available prey (e.g. megafaunal extinctions) as habitats shrunk with the expansion of humans and agriculture also suggest the plausibility of wolf population declines and lineage turnover. A remaining alternative for our inferred population phylogeny is that the basal lineage was absorbed into the three lineages sampled. Such a hypothesis is questionable though, as it requires there to be enough effective gene flow among each of the three wolf lineages such that no single lineage today serves best as a proxy for the basal lineage in our analysis.

We find consistent estimates for the timing of dog-wolf divergence across several models using *G-PhoCS* ($0.50 \times 10^{-4}$ substitution units, CI:0.46-0.53 $\times 10^{-4}$), which is very close to the divergence time between the wolf lineages (Fig. S10.4). This divergence time is consistent with recent estimates from shotgun sequences (~14 kya, CI:11-18 kya or 30 kya, CI:15-90 kya depending on assumed gene flow [51], using identical calibrations of $\mu = 1 \times 10^{-8}$/gen and 3 years/gen), but our results provide a considerably narrower credible interval (14-16kya, using the same calibrations). Given that dogs may have diverged from a basal wolf lineage not included in our sample, the estimated wolf-dog divergence time in our model is an upper bound for the most recent population-level divergence of dogs from wolves. We can also form a lower bound assuming a single origin of dogs by using the estimated divergence dates of the dog lineages. In our *G-PhoCS* results, the divergence among the Dingo and Boxer/Basenji ancestor takes place at $0.43 \times 10^{-4}$, (CI:$0.39 \times 10^{-4}$ – $0.46 \times 10^{-4}$) substitution units in the past and implies a lower bound for the wolf-dog divergence time between 11-14 kya, and then conservative bounds of 11-16 kya for the domestication event. Despite this narrow interval, the major source of uncertainty then lies in assumptions about mutation rates and generation times. For example, using an alterative mutation rate estimate of $2.2 \times 10^{-9}$/gen [52], (as done in one recent dog domestication study [20]), changes the bounds to 53 kya-72 kya. Notably, both mutation rate assumptions put the wolf-dog divergence prior to the origins of agriculture in humans.

The pre-agricultural origins of dogs we infer raises questions about the hypothesis that the advent of agriculture created a novel niche that was the driving force in dog domestication [15]. Here, we confirmed that amylase copy number expanded across almost all dog breeds as previously reported [15], and for the first time show that this CNV is present in wolf populations, suggesting that the initial duplication was likely a standing variant when the domestication process began. Notably, this site appears to be in low copy number in dog lineages that are not associated with agrarian societies (Dingo and Husky). Our findings imply that the *AMY2B* expansion likely evolved more recently with the development of large agriculturally based civilizations in the Middle East, Europe and Eastern Asia. Interestingly, our top selection hit, which shows reduced diversity across all dogs surveyed, indicates a different locus for dietary evolution early in dog domestication. This region is centered on the *CCRN4L* gene, an important regulator of metabolic phenotypes. Evolution at the locus may have facilitated shifts of lipid content in the diet of early dogs as foraging opportunities diverged from those characteristic of gray wolf kill sites. Interestingly, the locus might also have a pleiotropic effect on bone growth through its effects on cell fate and the well-established growth regulator *IGF1*.

Our results also provide additional insights into the genetic basis of the adaptations that occurred under domestication. Because we find evidence for a strong, recent wolf bottleneck, we expect that at the onset of dog domestication, there was substantially more genetic diversity for selection to act on than observed in modern wolves. Moreover, when we investigate outlier regions of the genome with respect to selection signatures, we found novel evidence for enrichment in gene categories involved in skeletal and dental morphology. Genes in these categories may have played roles in the evolution to early dogs having shortened, broader skulls, more extreme tooth crowding, smaller carnassials, and reduced body size [53]. Further, we also found evidence for selection on genes involved in neural development [7,14,15]. Notably, four of our top eight selection candidate regions each contain a gene known to impact memory and behavior in mice and humans. We also showed that many outlier regions have no known coding loci within them, and further that diversity is exceptionally reduced around dog-wolf differences in regulatory variant categories such as 5' and 3' UTR sites, as well as conserved non-coding elements. These results suggest that, as in other species [54-57], mutations at regulatory sites played a key role in adaptation. It is worth noting that some of the extreme selection signatures in this study may be false positives, in part due to the confounding effects of bottlenecks on selection scans [31,58], especially as we have shown each of the lineages sampled experienced strong bottlenecks in their history. Future sequencing studies in broader panels of dogs will help refine the most likely adaptive regions in dog domestication.

Overall, the genomes in this study reveal a dynamic and complex genetic history interrelating dogs and wolves. Post-divergence admixture is a key feature that complicates inferences of the origins of dogs, and the potential decline and turnover of wolf lineages may further complicate inference. Indeed, one interpretation of our results is that the lineage of wolves from which dogs were originally domesticated has gone extinct. If true, this hypothesis suggests ancient DNA studies will be crucial to substantially advance our understanding of the rapid transition from a large, aggressive carnivore to the omnivorous domestic companion that is a fixture of modern civilization.

## Materials and Methods

### Samples and Sequencing

We selected six samples for genome sequencing (Text S1). For all individuals besides the Chinese wolf, we used a combination of SOLiD (single end and long mate pair) and Illumina HiSeq paired end (PE) libraries, while for the Chinese wolf we only used Illumina PE data, as it was provided subsequent to our sequencing efforts for the other lineages (Text S2). For most downstream analyses, we also utilized sequence information from the Boxer reference genome (CanFam 3.0).

**Sequence Alignment, Genotyping, and Filters**

We aligned sequence reads to CanFam 3.0, with post-processing of aligned reads including the removal of duplicates, local realignment, and base quality recalibration (Text S3). We then genotyped each sample individually, using the Genome Analysis Toolkit (GATK) pipeline [59]. At the genome level we excluded repeats of recent origin, CpG sites, regions falling in copy number variants, and triallelic sites, while at the sample level we filtered out sites proximate to called indels, low quality genotypes, sites with excess depth of coverage, as well as all SNVs that were with 5 base pairs of another SNV (Text S4).

**Genotype Validation**

We compared genotype calls based upon sequencing (NGS) to those made for the same samples using the Illumina CanineHD BeadChip, which consists of >170,00 markers evenly spaced throughout the dog genome (Text S5). We also intersected genomic positions genotyped from sequencing in our samples, with genotypes generated using array/chip technology for a large panel of dogs and wolves, and performed PCA on the combined data set to verify that NGS genotypes clustered with array genotypes for the same lineages (Text S5).

**Structural Variant Detection**

We delineated segmental duplications in our six genomes by identifying regions with a significant excess depth of coverage (Text S6). For this purpose, we aligned Illumina and SOLiD reads with MrFAST [60] and drFAST [61] respectively. Absolute copy numbers were caluculated using mrCaNaVar version 0.31 (http://mrcanavar.sourceforge.net/). In the particular case of the previously reported amylase (*AMY2B*) expansion in the dog lineage [15] we also examined patterns of copy number across 52 breed dogs, six Dingoes, and 40 wolves using qPCR (Text S6).

**Functional Element Annotation**

All transcript level analyses were based upon gene annotations from the union of refGene, Ensembl and SeqGene annotation databases, with the condition that all annotated transcripts had proper start and stop codons, and contained no internal stop codons (Text S7, S8). In addition, we defined conserved non-coding elements (CNEs) on the basis of phastCons scores [62] (Text S7). Intersecting these annotations with our genome and sample-level filters, we defined sites that were fixed for different alleles between dogs and wild canids according to their functional class (hereafter dog sample fixed variants, DSFV). We then defined a more restricted subset of these loci as likely occurring at high frequency across all dogs, by filtering out sites where the dog-specific allele was at less that 75% frequency across a panel of an additional 12 dog breeds sequenced to low coverage (Text S11).

**$N_e$ Through Time**

We used the methods developed by Li and Durbin [22] to infer the trajectory of population sizes across time for the six canid genome sequences (Text S9). To translate from time units of generations to calendar years, we assume a generation time of three years for the wolves and the golden jackal. Following Lindblad-Toh et al. [18], the mutation rate assumed was 1.0 x $10^{-8}$ per generation.

**Testing for Admixture: ABBA-BABA**

To investigate the extent of gene flow between wolves and dogs subsequent to their divergence, we employed a method recently developed by Durand et al. [23]. This method tests for directional gene flow by testing for asymmetries in allele sharing between a source lineage (P3), and either of two receiving lineages (P1, P2) with reference to an outgroup (O). To focus on gene flow most germane to evolutionary processes influencing wolf-dog divergence, we restricted testing to those cases where when one of the dog samples was P3, the other two (P1 and P2) were wolves, and vice versa (P3=wolf, P1 and P2 =dogs). For more details, see Text S9.

**Demographic Model for Dog Domestication**

Our main demographic analysis is based on the Generalized Phylogenetic Coalescent Sampler (*G-PhoCS*) developed by Gronau et al. [21] and which we applied to 16,434 1kb loci chosen via a strict set of criteria to obtain putatively neutral loci (Text S10). The prior distributions over model parameters was defined by a product of Gamma distributions using the default setting chosen by Gronau [21]. Markov Chain was run for 100,000 burn-in iterations, after which parameter values were sampled for 200,000 iterations every 10 iterations, resulting in a total of 20,001 samples from the approximate posterior. Convergence was inspected manually for each run. We conditioned inference on the population phylogeny based upon the neighbor-joining tree constructed from the genome-wide distance matrix described above (Fig S9.1.1). We also constructed models under a 'multiple domestication' scenario, in which each dog lineage originated from a wolf lineage from the same geographic region, i.e. Basenji from Israeli wolf, Boxer from Croatian wolf, and Dingo from Chinese wolf. We assessed models in which the branch ancestral to dogs was sister to a particular extant wolf population, or one of internal branches in the wolf clade. In addition, we investigated the sensitivity of parameter estimates to choice of locus length, number of loci, intra-locus recombination, distance from coding exons, and selection on linked sites. For more details, see Text S10.

**Impact of Selective Sweeps**

To determine whether diversity is significantly reduced around functional sites compared to neutral baseline sites, we calculated the difference in mean dog nucleotide diversity between neutral and functional DSFV sites (at overall high frequency in dogs as described above), in non-overlapping 10kb windows across a 2Mb interval centered on those sites, using the following test statistic:

$$\pi_{N-F} = \sum_{i=1}^{n} \left( \ln[\hat{\pi}_{i,neutral}] - \ln[\hat{\pi}_{i,functional}] \right),$$

where $i$ is an index reflecting the relative position of the window over the 2Mb genomic interval, $n$ equals the total number of windows in that interval, and $\hat{\pi}_i$ represents the mean nucleotide diversity at that relative position. Neutral baseline sites were defined as noncoding DSFV sites at high frequency in dogs that were within 5kb of genes that did not contain DSFV sites, with this latter criterion employed in order to exclude sites with reduced diversity due to selective sweeps occurring at linked sites within genic regions. Significance testing was carried out via permutation tests, with 5,000 permutations per test. For more details, see Text S12.

**Genome-wide Selection Scans**

To identify regions of the genome bearing signatures of positive selection, we computed summary statistics in 100kb sliding windows across the genome, incremented in steps of 10kb. Within each window, we compute three selection scan statistics for sites passing our genome and sample-level filters: ratio of nucleotide diversities ($\Delta\pi$) $=\pi_{wolf}/\pi_{dog}$, $F_{ST}$, and the diifference in Tajima's D ($\Delta$TD)=$TD_{wolf}$ -$TD_{dog}$. We excluded windows with <30,000 pass-filter sites, and identified outlier windows by calculating a joint empirical (product) percentile of the three statistics in each window, and ranking windows by this joint statistic. Windows were collapsed into regions when a pair of windows fell within 200kb of each other, and outlier regions were identified according to the ranking of the maximum joint percentile statistic computed for a window within the region. Intersecting the outlier regions with our gene annotations, we then tested for enrichments in in Gene Ontology (GO) categories, Kegg/Reactome pathways (KGR) and Human Phenotype Ontologies (HPO). For more details, see Text S13.

**ACKNOWLEDGMENTS.** We thank B. Chin, T. Toy, Z. Chen and the UCLA DNA Microarray Facility for library preparations and sequencing done at UCLA; D. Wegmann for initial development of the *VcfAnnotator* program used in analyses here; A. Platt for feedback on analyses and manuscript; R. Hefner and The National Collections of Natural History at Tel Aviv University for procuring and access to samples. We thank E. Randi, R. Godinho, and B. Yue for facilitating visits of MG, PMS, and ZF to the lab of RKW. This work was supported by an NSF Postdoctoral Fellowship DBI-0905784 (AHF), NSF Graduate Research Fellowship (RMS), Searle Foundation Scholar Award (JN), NSF grant EF-1021397 (JN, RKW, AHF, RMS, EH), NIH T32 HG002536 (EH), NIH (NIGMS) grant GM102192 (IG), UC MEXUS-CONACYT doctoral fellowship 213627 (DOD), PhD grant from Fundação para a Ciência e a Tecnologia (Portugal) SFRH/BD/60549/2009 (PMS), PhD Grant from University of Bologna (Italy), XXIV cicle, Biodiversity and Evolution (MG), National Science and Technology Support Project of China 2012BAC01B06 (ZF), Rosztoczy Foundation (PM), USHHS Ruth L. Kirschstein Institutional National Research Service Award #T32 CA009056 (KS), National Institute of Bioinformatics (BL-G), Intramural Program of the National Human Genome Research Institute (HB, EAO), JAEDOC-CSIC (EFS) (OR), Marie Curie CIG PCIG10-GA-2011-303772 (CA) Programa de Captación del Conocimiento para Andalucía for sequencing of Chinese wolf at BGI, ERC Starting Grant 260372 and MICINN (Spain) BFU2011-28549 (TM-B), and NSF grant DEB-0948510 (ARB and CDB.).

**Figure 1**. **Sampling, heterozygosity, comparison of next generation sequencing with array typed samples, and historical changes in effective population size.** (A) Geographic distribution of sampled lineages. (B) Box plots of heterozygosity measured in 5000 100kb windows for each sample. (C) PCA plot of next-generation sequencing

(NGS) samples generated in this study (open circles) along with corresponding samples genotyped on the Affymetrix canid array [7] (two letter codes: M=Mid-East Wolf, E=European Wolf, Ch=Chinese Wolf, Ba=Basenji, Bo=Boxer, D=Dingo, J=Golden Jackal). (D) Reconstruction of historical patterns of effective population size ($N_e$) for individual genome sequences. Based upon the genomic distribution of heterozygous sites using the pairwise sequential Markovian coalescent (PSMC) method of Li and Durbin 2011[22]. Time scale on the x-axis is calculated assuming a mutation rate of $1 \times 10^{-8}$ per generation (see SI Text S9.2); estimates from the full data and 50 bootstraps are depicted by darker and lighter lines, respectively.

**Figure 2. Demographic model of domestication.** Divergence times, effective population sizes ($N_e$), and post-divergence gene flow inferred by *G-PhoCS* in joint analysis of the boxer reference genome, and the sequenced genomes of two basal dog breeds, three wolves, and a golden jackal. The width of each population branch indicates inferred population size. The width of the outer grey area denotes the upper edge of the 95% Bayesian credible interval, and the inner colored band indicates the lower edge of the interval. Vertical error bars (gray) indicate 95% Bayesian credible intervals for estimated divergence times. Arrows indicate migration bands along which significant gene flow was inferred, with the vertical size of the dark/gray arrowhead indicating the lower and upper credible intervals on the magnitude of gene flow, such that an arrowhead with a vertical height of 1 on the y-axis corresponds to migration rate of m=0.5. Panels show parameter estimates for (A) the population tree best supported by genome-wide sequence divergence (Fig. S9.1.1), (B) a multiple domestication model, and (C) a single wolf lineage origin model in which dogs diverged most recently from the Israeli wolf lineage (similar star-like divergences are found assuming alternative choices for the single wolf ancestor (see SI text S10). Estimated divergence times and effective population sizes are calibrated assuming an average mutation rate of $1 \times 10^{-8}$ substitutions per generation and an average generation time of three years. See SI Text S10, including Fig. S10.4 and Table S10.2.

**Figure 3. Signals of selective sweeps.** Reduced nucleotide diversity due to selective sweeps around sites with dog-specific high-frequency/fixed derived alleles. The non-coding (dashed black-line) serves as a reference for the reduction expected around putatively neutral substitutions. CNE=conserved non-coding elements. See SI Text S11 for details.

**Figure 4. Regions under selection.** (A) Summary statistics for the top selection scan hit region, centered on CCRN4L. The outlier region is shown in grey, empirical p-values of three summary statistics and the composite empirical p-values are shown, along with dog-specific variants (DSFV, see Text S11) found in the region in the top plot. Genes annotated as unknown are indicated with an *. The middle plots show two measures of genetic diversity ($\theta_\pi^{dog}$, $\theta_\pi^{wolf}$, solid lines) and Tajima's D statistics (dashed lines), with blue for the dogs and red for the wolves. The bottom plot represents the genotypes observed in six individuals in the same region. (B) Top 20 outlier regions ranked by joint percentile of selection scan statistics**.** Columns within "This study" are based on the sequencing data generated here, while those under CanMap are computed from a ~48k SNP data set for a large set of wolves and ancient/basal dog breeds (SI Text S13). Heat

map colors reflect upper percentiles of the calculated metrics, with warmer colors indicating higher percentiles. Overlaps with previous studies: 1, vonHoldt *et al.* 2010 [7]; 2, Axelsson *et al.* 2013 [15], with numbers indicating region ids. No overlaps were observed with Boyko *et al.* 2010 [11] or Vaysse *et al.* 2011 [63]. For overlaps with lower ranked regions and for fixed-sized windows, see SI Text S13, S14.

**Figure 5. Copy number variation at amylase (*AMY2B*) locus.** (A) Copy number variation (CNV) at *AMY2B* estimated from whole genome sequence data, showing presence of elevated copy number in Basenji but not in other lineages. Results are based on SOLiD data, except for the Chinese wolf (see SI Text S6.2 for supporting results and SI Text S15 for CNV analyses in an additional 12 dog breeds). (B) qPCR results on CNV state in an expanded set of wolf and dog lineages. Abbreviations for lineages are: AFG, Afgan Hound; AFR, Africanis; AKI, Akita; BSJ, Basenji; BE, Beagle; BU, Bulldog, CAN, Canaan Dog; CU, Chihuahua; CC, Chinese Crested; FC, Flat-coated Retriever; GD, Great Dane; IH, Ibizan Hound; KUV, Kuvasz; MAS, Mastiff; NGS, New Guinea Singing Dog; PEK, Pekinese; PHU, Phu Quoc; SAL, Saluki; SAM, Samoyed; SCT, Scottish Terrier; SHA, Shar Pei; SIH, Siberian Husky; THD, Thai Dog; TOP, Toy Poodle; DNG, Dingo; CHW, Chinese wolf; INW, Indian wolf; ISW, Israeli wolf; ITW, Italian wolf; RUW, Russian wolf; SPW, Spanish wolf; YSW, Yellowstone wolf; GLW, Great Lakes wolf.

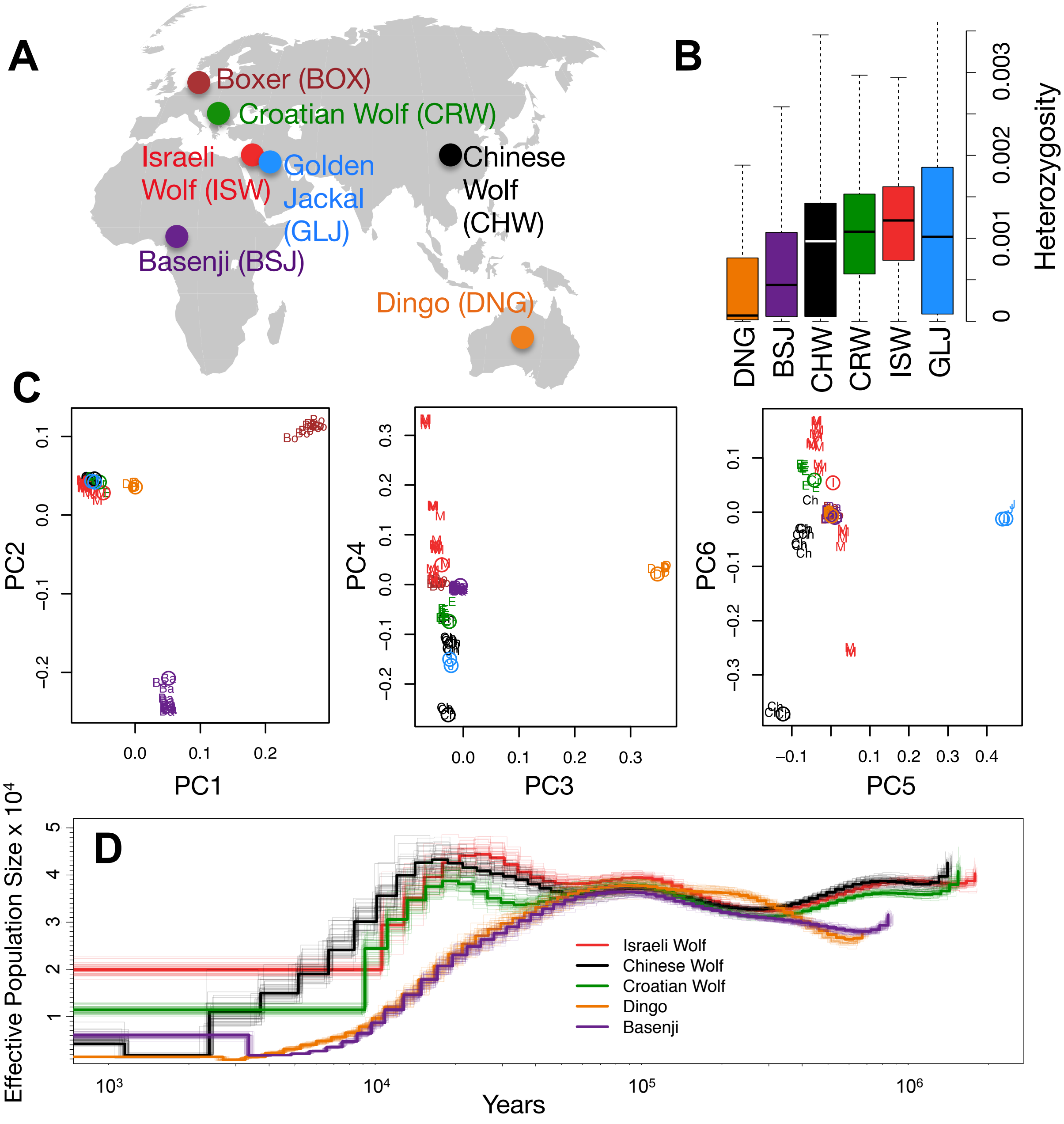

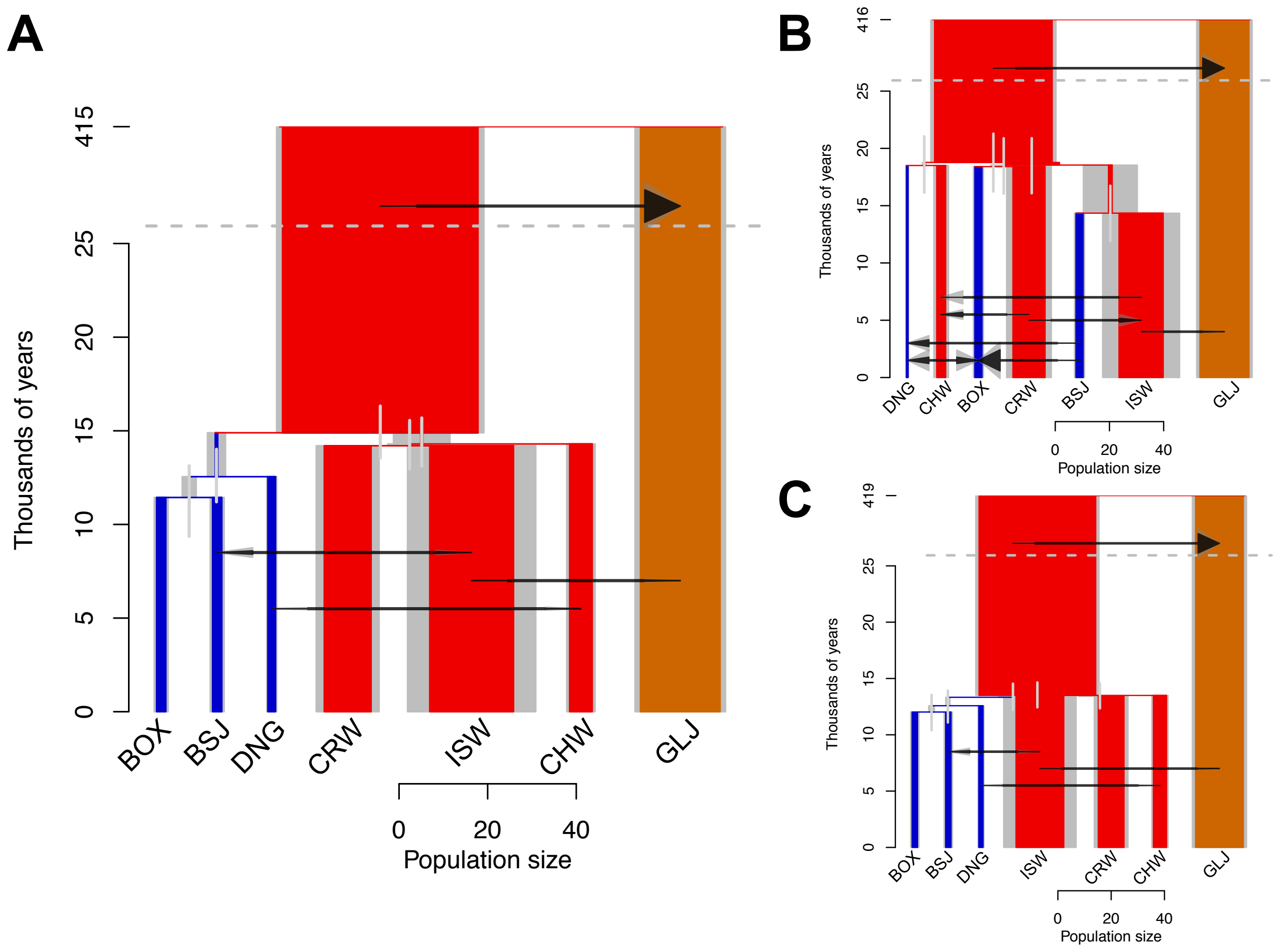

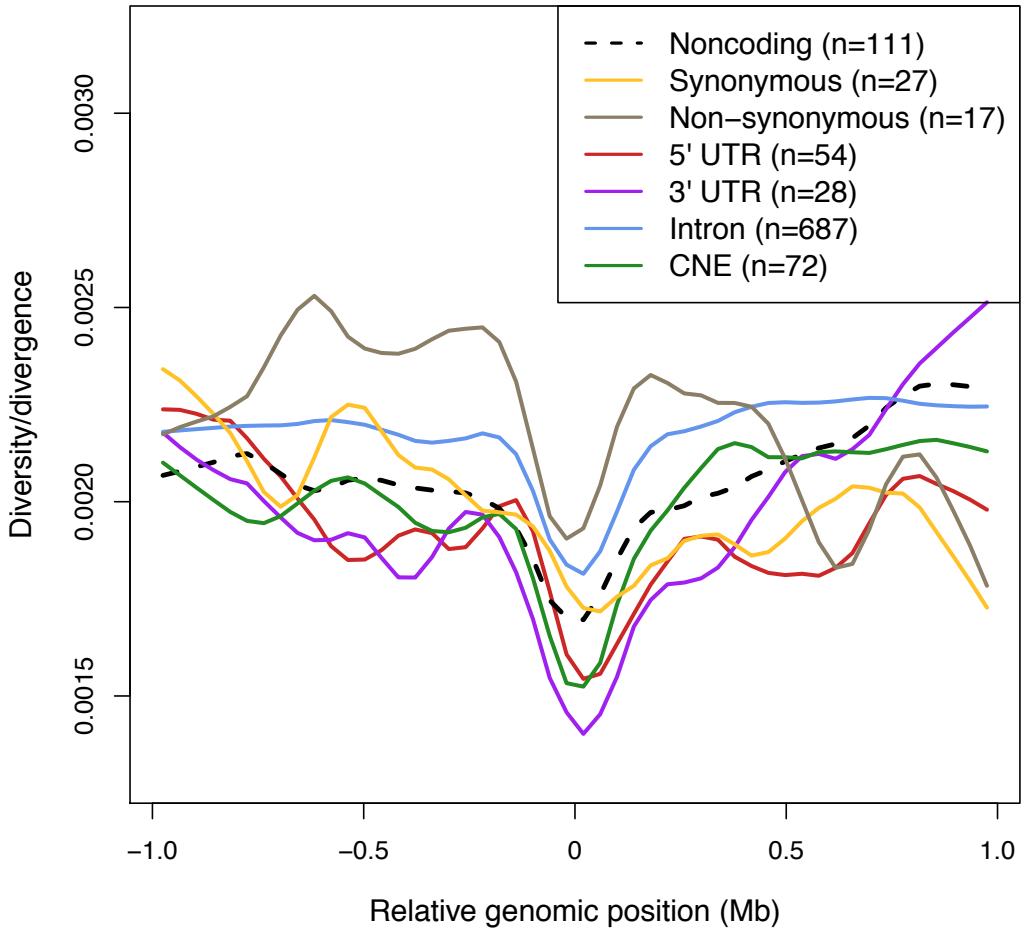

A

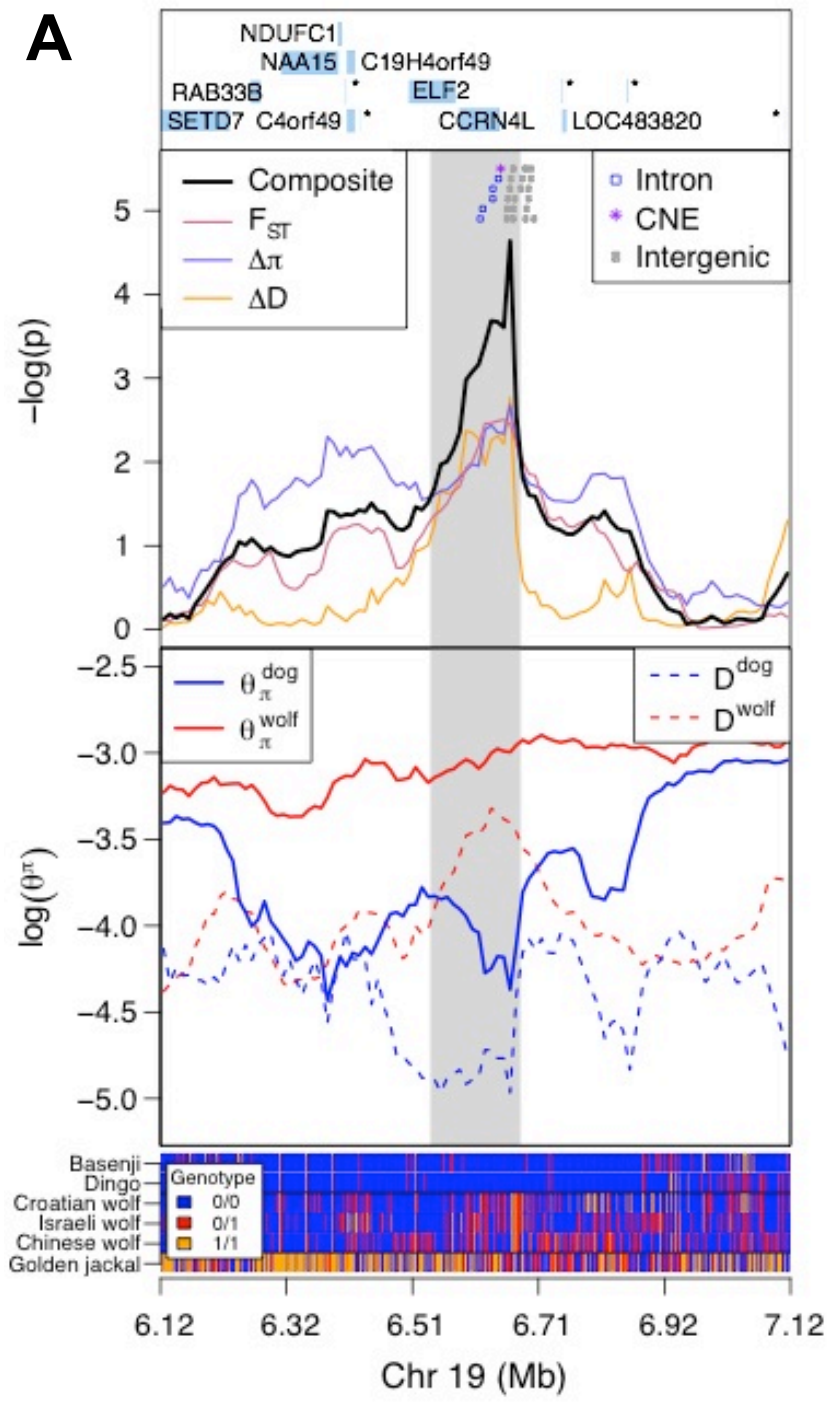

B

| Region | This Study | | | | | CanMap | | | Previous Studies | | Genes |
|---|---|---|---|---|---|---|---|---|---|---|---|
| | Δπ | Fst | ΔTD | Joint% | #Outliers | Δπ | Fst | Joint% | 1 | 2 | |
| 19:6.5–6.73 | 2.68 | 2.51 | 2.77 | 5.24 | 14 | 0.58 | 0.83 | 97.2 | 93.1 | | ELF2,CCRN4L |
| 18:18.55–18.77 | 3.44 | 2.59 | 2.34 | 4.64 | 13 | 0.59 | 0.70 | 95.8 | 94.8 | | LHFPL3 |
| 10:6.65–7.26 | 4.76 | 2.68 | 3.00 | 4.54 | 31 | 0.30 | 0.61 | 92.0 | | 18 | LOC481139,EN28388 |
| 31:4.95–5.11 | 2.45 | 2.65 | 2.91 | 4.29 | 7 | 0.54 | 0.40 | 87.4 | | | CADM2 |
| 2:24.93–25.16 | 1.89 | 4.54 | 2.90 | 4.09 | 14 | 0.18 | 0.19 | 50.4 | | | LOC607279,FAM107B |
| 11:40.6–40.89 | 2.70 | 3.24 | 2.17 | 4.04 | 18 | −0.80 | 0.47 | 18.9 | | 19 | SH3GL2,EN01567 |
| 2:49.34–49.58 | 2.76 | 1.80 | 3.28 | 3.94 | 14 | −1.39 | 0.34 | 8.4 | | | PDE4D,cfa−mir−582 |
| 26:11.67–11.86 | 2.47 | 2.11 | 2.45 | 3.92 | 10 | −0.05 | 0.10 | 14.7 | | | CUX2,EN27618 |
| 5:42.07–42.21 | 2.64 | 2.65 | 1.73 | 3.82 | 5 | −0.30 | 0.57 | 37.4 | | | FAM18B2,EN17928,XM_858503.1, TRIM16,EN27757,ZNF79,LOC608913 |
| 1:31.31–31.49 | 1.98 | 2.63 | 2.16 | 3.72 | 9 | 0.10 | 0.51 | 78.7 | 93.3 | | |
| 12:62.03–62.16 | 3.33 | 2.76 | 1.62 | 3.70 | 4 | −1.06 | 0.01 | 1.75 | | | LOC475009,EN03564,EN27834 |
| 24:7.32–7.5 | 1.86 | 2.28 | 2.58 | 3.68 | 9 | 0.12 | 0.12 | 33.9 | | | C20orf79,C24H20orf79,DTD1 |
| 21:7.6–7.81 | 2.69 | 1.71 | 2.97 | 3.63 | 12 | 0.01 | 0.47 | 65.4 | | | LOC485113,JRKL,CCDC82 |
| 18:26.99–27.62 | 3.00 | 1.88 | 2.6 | 3.62 | 36 | 0.30 | 0.55 | 90.6 | 90.3 | | EN25217,LOC610706,LOC610718, SEMA3D,EN27671,LOC610738,EN28042 |
| 37:9.53–9.72 | 2.45 | 2.54 | 2.17 | 3.59 | 7 | 0.28 | 0.32 | 72.7 | | | ANKRD44 |
| 31:6.03–6.22 | 2.41 | 1.68 | 2.40 | 3.58 | 10 | | | | 93.9 | | |
| 6:31.33–31.52 | 2.18 | 1.90 | 2.24 | 3.56 | 10 | 0.24 | 0.52 | 87.1 | | | C6H16orf45,MPV17L,PDXDC1,EN26265, NTAN1,RRN3 |
| 11:32.56–32.88 | 1.83 | 2.05 | 2.09 | 3.54 | 9 | 1.26 | 0.77 | 99.3 | | | EN27693 |
| 1:42.04–42.21 | 1.95 | 2.26 | 1.87 | 3.53 | 8 | 0.45 | 0.29 | 74.8 | | | |
| 5:6.88–7.29 | 2.45 | 1.87 | 2.39 | 3.49 | 27 | 0.35 | 0.76 | 93.7 | 93.7 | | EN20853,SNX19,EN23101 |

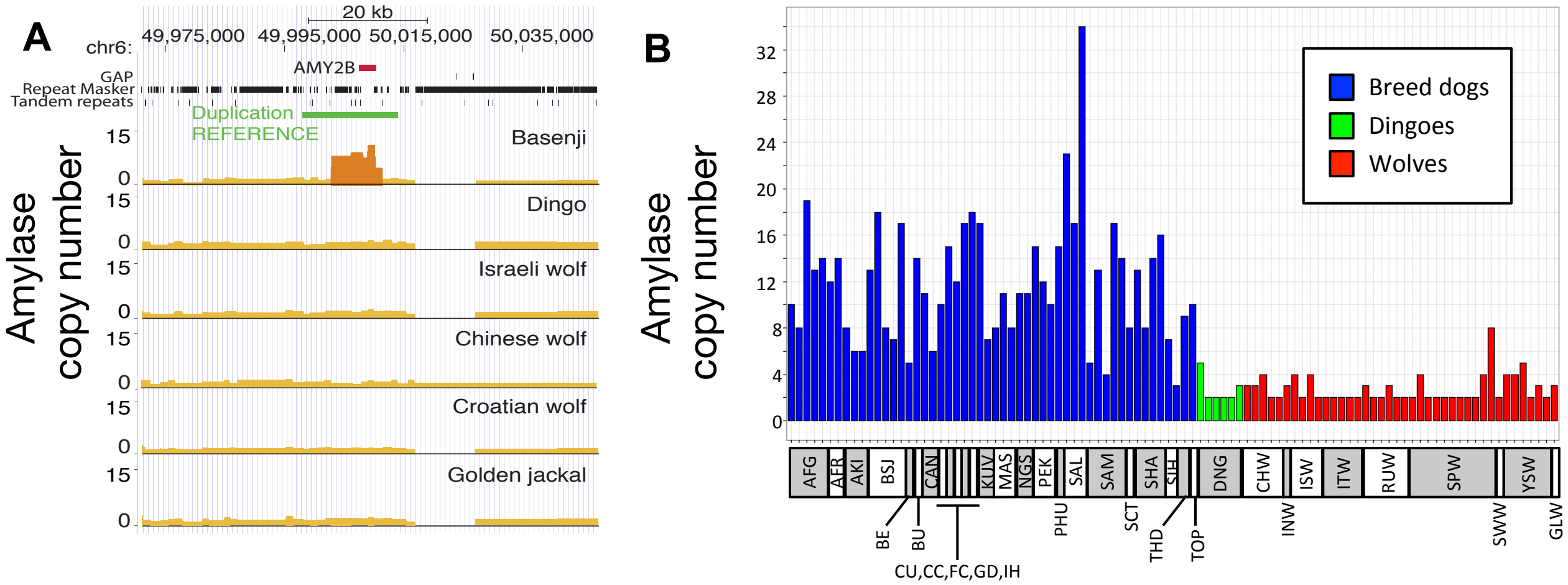